\documentclass[letter, traditabstract, longauth]{aa} 
\usepackage{graphicx}
\usepackage{txfonts}
\usepackage[dvips]{color}
\usepackage{sidecap}
\usepackage{latexsym,natbib}
\def\cf{cf.~}

\def\fig{Fig.\,}
\def\sec{Sect.\,}
\def\hi{$\mathrm{H}$\,{\sc i}\,}
\def\hii{$\mathrm{H}_2$\,}
\def\ltsim{~\rlap{\lower -0.5ex\hbox{$<$}}{\lower 0.5ex\hbox{$\sim\,$}}}
\def\gtsim{~\rlap{\lower -0.5ex\hbox{$>$}}{\lower 0.5ex\hbox{$\sim\,$}}}
\def\arcsec{$^{\prime\prime}$}
\def\arcmin{$^{\prime}$}
\begin{document}
%
\title{Radial distribution of gas and dust in spiral galaxies\thanks{Herschel is an ESA space observatory with science
  instruments provided by European-led Principal Investigator consortia and
  with important participation from NASA}}
\subtitle{The case of M\,99 (NGC\,4254) and M\,100 (NGC\,4321)}

\author{
	  M. Pohlen\inst{1},
	  L. Cortese\inst{1},
	  M. W. L. Smith\inst{1},
	  S. A. Eales\inst{1},
          A. Boselli\inst{3},
	  G. J. Bendo\inst{2},
	  H. L. Gomez\inst{1},
 	  A. Papageorgiou\inst{1},
	  R. Auld\inst{1},
	  M. Baes\inst{4},
	  J. J. Bock\inst{5},
	  M. Bradford\inst{5},
	  V. Buat\inst{3},
	  N. Castro-Rodriguez\inst{6},
	  P. Chanial\inst{7},
	  S. Charlot\inst{8},
	  L. Ciesla\inst{3},
	  D. L. Clements\inst{2},
          A. Cooray\inst{9},
	  D. Cormier\inst{7},
	  E. Dwek\inst{10},
	  S. A. Eales\inst{1},
	  D. Elbaz\inst{7},
	  M. Galametz\inst{7},
	  F. Galliano\inst{7},
	  W. K. Gear\inst{1},
          J. Glenn\inst{11},
	  M. Griffin\inst{1},
	  S. Hony\inst{7},
	  K. G. Isaak\inst{1,12},
	  L. R. Levenson\inst{5},
	  N. Lu\inst{5},
	  S. Madden\inst{7},
	  B. O'Halloran\inst{2},
	  K. Okumura\inst{7},
	  S. Oliver\inst{13},
	  M. J. Page\inst{14},
          P. Panuzzo\inst{7},
	  T. J. Parkin\inst{15},
	  I. Perez-Fournon\inst{6},
	  N. Rangwala\inst{11},
	  E. E. Rigby\inst{16},
	  H. Roussel\inst{8},
	  A. Rykala\inst{1},
	  N. Sacchi\inst{17},
	  M. Sauvage\inst{7},
	  B. Schulz\inst{18},
	  M. R. P. Schirm\inst{15},
	  M. W. L. Smith\inst{1},
	  L. Spinoglio\inst{17},
	  J. A. Stevens\inst{19},
	  S. Srinivasan\inst{8},
	  M. Symeonidis\inst{14},
	  M. Trichas\inst{2},
	  M. Vaccari\inst{20},
	  L. Vigroux\inst{8},
	  C. D. Wilson\inst{15},
	  H. Wozniak\inst{21},
	  G. S. Wright\inst{22},
	  W. W. Zeilinger\inst{23}
          }

\institute{	
	School of Physics and Astronomy, Cardiff University, Queens  
Buildings The Parade, Cardiff CF24 3AA, UK
	 \and
		Astrophysics Group, Imperial College, Blackett Laboratory, Prince  
Consort Road, London SW7 2AZ, UK
	 \and
		Laboratoire d'Astrophysique de Marseille, UMR6110 CNRS, 38 rue F.  
Joliot-Curie, F-13388 Marseille France
          \and
		Sterrenkundig Observatorium, Universiteit Gent, Krijgslaan 281 S9,  
B-9000 Gent, Belgium
	 \and
		Jet Propulsion Laboratory, Pasadena, CA 91109, United States;  
Department of Astronomy, California Institute of Technology, Pasadena,  
CA 91125, USA
	\and
		Instituto de Astrof\'isica de Canarias, v\'ia L\'actea S/N, E-38200 La  
Laguna, Spain
	\and
CEA, Laboratoire AIM, Irfu/SAp, Orme des Merisiers, F-91191
Gif-sur-Yvette, France
	\and
		Institut d'Astrophysique de Paris, UMR7095 CNRS, Universit\'e Pierre  
\& Marie Curie, 98 bis Boulevard Arago, F-75014 Paris, France
\and
Department of Physics \& Astronomy, University of California, Irvine,
CA 92697, USA 
              \and	
		Observational  Cosmology Lab, Code 665, NASA Goddard Space Flight   
Center Greenbelt, MD 20771, USA
	\and
		Department of Astrophysical and Planetary Sciences, CASA CB-389,  
University of Colorado, Boulder, CO 80309, USA
\and
ESA Astrophysics Missions Division, ESTEC, PO Box 299, 2200 AG
		Noordwijk, The Netherlands
	\and
		Astronomy Centre, Department of Physics and Astronomy, University of  
Sussex, UK
	\and
		Mullard Space Science Laboratory, University College London,  
Holmbury St Mary, Dorking, Surrey RH5 6NT, UK
	\and
		Dept. of Physics \& Astronomy, McMaster University, Hamilton,  
Ontario, L8S 4M1, Canada
	\and
		School of Physics \& Astronomy, University of Nottingham, University  
Park, Nottingham NG7 2RD, UK
	\and
		Istituto di Fisica dello Spazio Interplanetario, INAF, Via del Fosso  
del Cavaliere 100, I-00133 Roma, Italy
	\and
		Infrared Processing and Analysis Center, California Institute of  
Technology, Mail Code 100-22, 770 South Wilson Av, Pasadena, CA 91125,  
USA
	\and
		Centre for Astrophysics Research, Science and Technology Research  
Centre, University of Hertfordshire, College Lane, Herts AL10 9AB, UK
	\and
		University of Padova, Department of Astronomy, Vicolo Osservatorio  
3, I-35122 Padova, Italy
	\and
		Observatoire Astronomique de Strasbourg, UMR 7550 Universit\'e de  
Strasbourg - CNRS, 11, rue de l'Universit\'e, F-67000 Strasbourg
\and
UK Astronomy Technology Center, Royal Observatory Edinburgh, Edinburgh, EH9 3HJ, UK 
	\and
		Institut f\"ur Astronomie, Universit\"at Wien, T\"urkenschanzstr. 17,  
A-1180 Wien, Austria
}

   \date{Received ; accepted }

 
\abstract{ 

By combining {\it Herschel}-SPIRE data with archival {\it Spitzer}, \hi, and
CO maps, we investigate the spatial distribution of gas and dust in the two
famous grand-design spirals M\,99 and M\,100 in the Virgo cluster. Thanks to
the unique resolution and sensitivity of the {\it Herschel}-SPIRE photometer,
we are for the first time able to measure the distribution and extent of cool,
submillimetre (submm)-emitting dust inside and beyond the optical radius. We
compare this with the radial variation in both the gas mass and the
metallicity. Although we adopt a model-independent, phenomenological approach,
our analysis provides important insights. We find the dust extending to at
least the optical radius of the galaxy and showing breaks in its radial
profiles at similar positions as the stellar distribution. The colour indices
$f350/f500$ and $f250/f350$ decrease radially consistent with the temperature
decreasing with radius. We also find evidence of an increasing gas to dust
ratio with radius in the outer regions of both galaxies. }


\authorrunning{Pohlen et al.}
\titlerunning{Radial distribution of gas and dust in M\,99 and M\,100.}

   \maketitle
%

\section{Introduction}

\begin{figure*}
\hspace*{1cm}
SDSS u-z \hspace{1.5cm} 70 $\mu$m \hspace{2.1cm} 250 $\mu$m \hspace{1.9cm} 350 $\mu$m \hspace{2.0cm}
500$\mu$m \hspace{2.3cm} \hi \\[-0.15cm]
\includegraphics[height=18.5cm,angle=270]{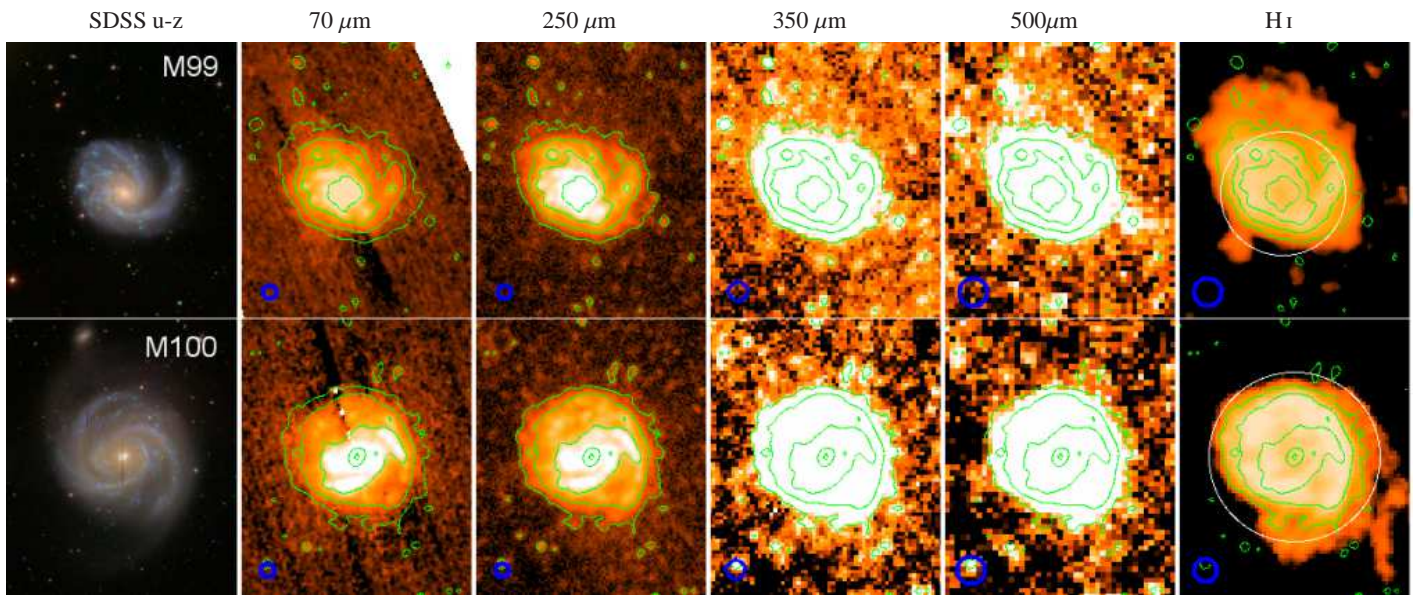}
\vspace*{-0.0cm}
\caption{The matched SDSS, MIPS 70 $\mu$m, SPIRE 250, 350, 500 $\mu$m, and \hi-maps
  of M\,99 {\it (upper row)} and M\,100 {\it (lower row)}. The {\it green
    contours} are SPIRE 250 at levels about 0.03($\approx\!5\sigma$), 0.10,
  0.34, 1.15, 3.9 Jy beam$^{-1}$. The 350 and 500 $\mu$m are displayed at a
  high contrast to show the extent of the dust emission and the
    scan-artefact free smooth background dominated only by confusion
    noise. The {\it white circle} on the \hi-maps marks the optical radius
  defined by $D25$. The {\it blue circles} on each map indicate the
  FWHM. Note, the SDSS jpg is scaled but not WCS matched.}
\label{maps}
\vspace*{-0.3cm}
\end{figure*}

The study of the gas and dust distribution in galaxies is essential to
understanding their formation and evolution. The rate at which gas is accreted
and converted into stars regulates not only the star formation history of
galaxies but also their chemical evolution. Dust is supposed to play a key
role in this process. Dust grains are the main coolant in star-forming
galaxies, shielding the gas from the UV radiation and representing the site at
which \hi\ is converted into \hii, and then collapses into stars \citep[see
  e.g., reviews by][]{Calzetti2001,Draine2003}.

To understand the dust, we need to map the cold component, which does not
dominate the energy but dominates the mass. However, using previously existing
facilities, our knowledge of the interplay between gas and dust and their
radial distribution have remained highly uncertain, being based only on
$\lambda\!<\!  160\mu$m space observations \citep[e.g.~][]{Munoz2009a,
  Bendo2010a}, and challenging ground-based submm observations that were of
optimal quality only at significantly longer wavelength. For many galaxies,
only integrated quantities have been derived because of the poor resolution of
previous satellites. Resolved submm studies from the ground remain limited to
the very nearby universe and large surveys of more distant galaxies are
unfeasible.

The SPIRE instrument \citep{Griffin2010} on-board {\sl Herschel}
\citep{Pilbratt2010} now bridges this gap observing in the range of
250$\mu$m-500$\mu$m. With the benefit of the stable conditions of a space
observatory, it is much more sensitive to the cold component and provides
excellent maps at high resolution, so is ideal for large surveys of many
galaxies. With {\sl Herschel}, we are now able to tackle the problem of the
interplay between gas and dust, and combined with the number of recent high
resolution surveys tracing the gas mass of local galaxies
\citep[e.g.~][]{Chung2009,Kuno2007}, we can study the distribution of gas,
metals, and dust on a kpc scale for hundreds of galaxies.

Here, we discuss first results for two famous grand-design spirals M\,99 and
M\,100 (see \fig\ref{maps}) in the Virgo cluster. We explore the distribution
of the cool dust traced with {\sl Herschel} by inspecting their radial profiles from
mid-infrared to submm wavelengths. We then attempt to correlate this with the
observed gas and metallicity distributions and search for temperature
variations. We use the {\sl Herschel}-SPIRE maps taken during {\sl Herschel}'s Science
Demonstration Phase. The two galaxies are part of the {\sl Herschel} Reference
Survey \citep{Boselli2010a}. This guaranteed time key project will provide
maps for a statistically-complete sample of 323 nearby galaxies in all three
SPIRE bands. In the RC3 \citep{rc3}, the classification and optical radius
$R25$ of M\,99 and M\,100 is given as SA(s)c, 2.69\arcmin ($\approx 12.9$kpc),
and SAB(s)bc, 3.71\arcmin($\approx 17.8$kpc), respectively. We assume a
distance for the Virgo Cluster of 16.5 Mpc \citep{Mei2007}.


\section{Data}

\subsection{{\sl Herschel}-SPIRE} 
The SPIRE photometer \citep{Griffin2008,Griffin2010} data were processed up to
Level-1 (i.e., to the level where the pointed photometer time-lines were
derived) with a customised pipeline script adapted from the official
pipeline ({\sl POF5\_pipeline.py}, dated 27 Nov 2009) provided by the SPIRE
Instrument Control Centre (ICC)\footnote{See 'The SPIRE Analogue Signal Chain
  and Photometer Detector Data Processing Pipeline' \citep{Griffin2009} for a
  more detailed description of the pipeline and a list of the individual
  modules.}.  This Jython script was run in the {\sl Herschel} Interactive Processing
Environment \citep[HIPE][]{Ott2010} coming with continuous integration build
number 3.0.327, which is the current developer's branch of the data
reduction software. However, in terms of the SPIRE scan-map pipeline up to
Level-1, this is in principle identical to the {\sl Herschel} Common Science
System/Standard Product Generation v2.1, even down to the
calibration files{\footnote{Aprt from the BsmPos file, for which we use an
    updated version that should improve the absolute astrometry.}  associated
  with the individual pipe\-line modules. This version of the pipeline is used
  at ESA to produce the standard products that will be available from the
  {\sl Herschel} Science Archive once they become public.

Currently, the Level-1 photometer time-lines still requires a residual
baseline subtraction to be made. However, instead of subtracting the median of
the time-line for each bolometer per scanleg (the default), we subtracted the
median of the time-lines for each bolometer over the whole observation. This
circumvents shadow artefacts caused in cases where the signal time-lines in
individual scanlegs are dominated by structured emission e.g. a large, extended
galaxy or a strong cirrus component.

This baseline-subtracted Level-1 data were then fed through an iterative
de-striper, which minimises the difference between the signal in individual
detector time-lines and the final map \citep[see][for a longer
  description]{Bendo2010b}.  At the end of this process, the signal time-lines
were then mapped into a final image using the Naive Mapper available in HIPE.

For M\,99, the de-striping approach left some residual large-scale gradients. In
this case, we resorted back to an initial baseline subtraction on a scan
by scan basis. However, instead of a median, we used a robust linear fit with
outlier rejection to the first and last fifty sample points, thus avoiding the
galaxies in the centre of the time-lines.

According to the ICC, the uncertainty in the flux calibration is of the order
of 15\% \citep{Swinyard2010} and is currently based on a preliminary
calibration.  However, the ICC has released some interim small correction
factors to improve this calibration. All flux values derived using the current
standard calibration file for the flux conversion, are multiplied by 1.02,
1.05, and 0.94, for the 250$\mu$m, 350$\mu$m, and 500$\mu$m,
respectively.\footnote{\tiny See
  http://herschel.esac.esa.int/SDP$\_$wkshops/presentations/IR/\\3$\_$Griffin$\_$SPIRE$\_$SDP2009.pdf.}
The full widths at half maximum (FWHM) of the SPIRE beams are 18.1\arcsec,
25.2\arcsec, and 36.9\arcsec, the pixel sizes are 6\arcsec, 10\arcsec, and
14\arcsec\ at 250, 350, and 500 $\mu$m, respectively.  

For both, M\,99 and M\,100, we observed a 12\arcmin $\times 12$\arcmin\ field
doing three repetitions of a cross-linked scan-map at nominal detector
settings and nominal scan speed (30\arcsec/sec). The M\,100 observation was
carried out twice. Both were treated independently here and used to verify the
consistency of our results.

\subsection{MIPS, CO, and HI}
 
\begin{figure*}
\begin{center}
\includegraphics[width=5.7cm,angle=0]{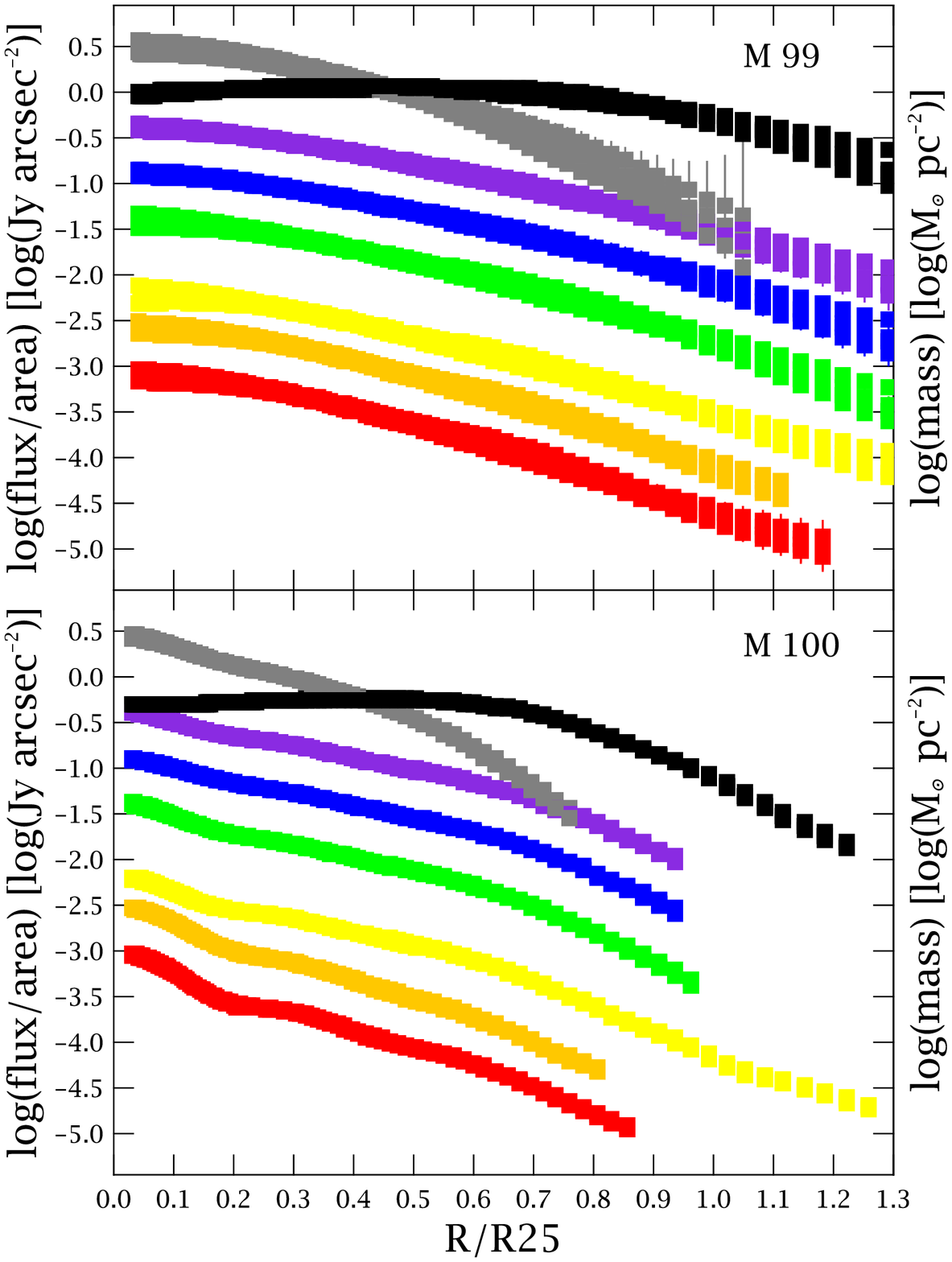}
\includegraphics[width=5.3cm,angle=0]{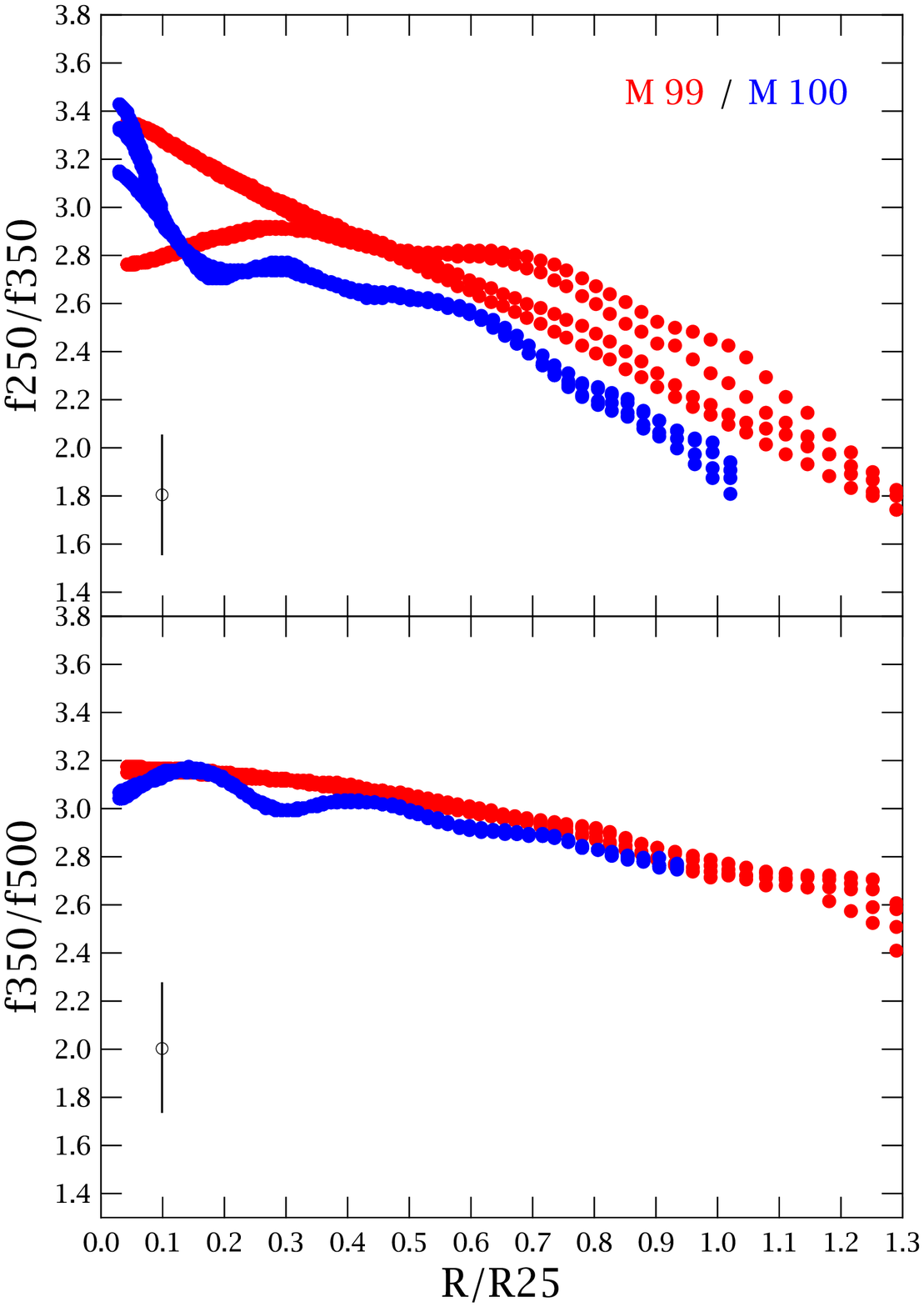}
\includegraphics[width=5.3cm,angle=0]{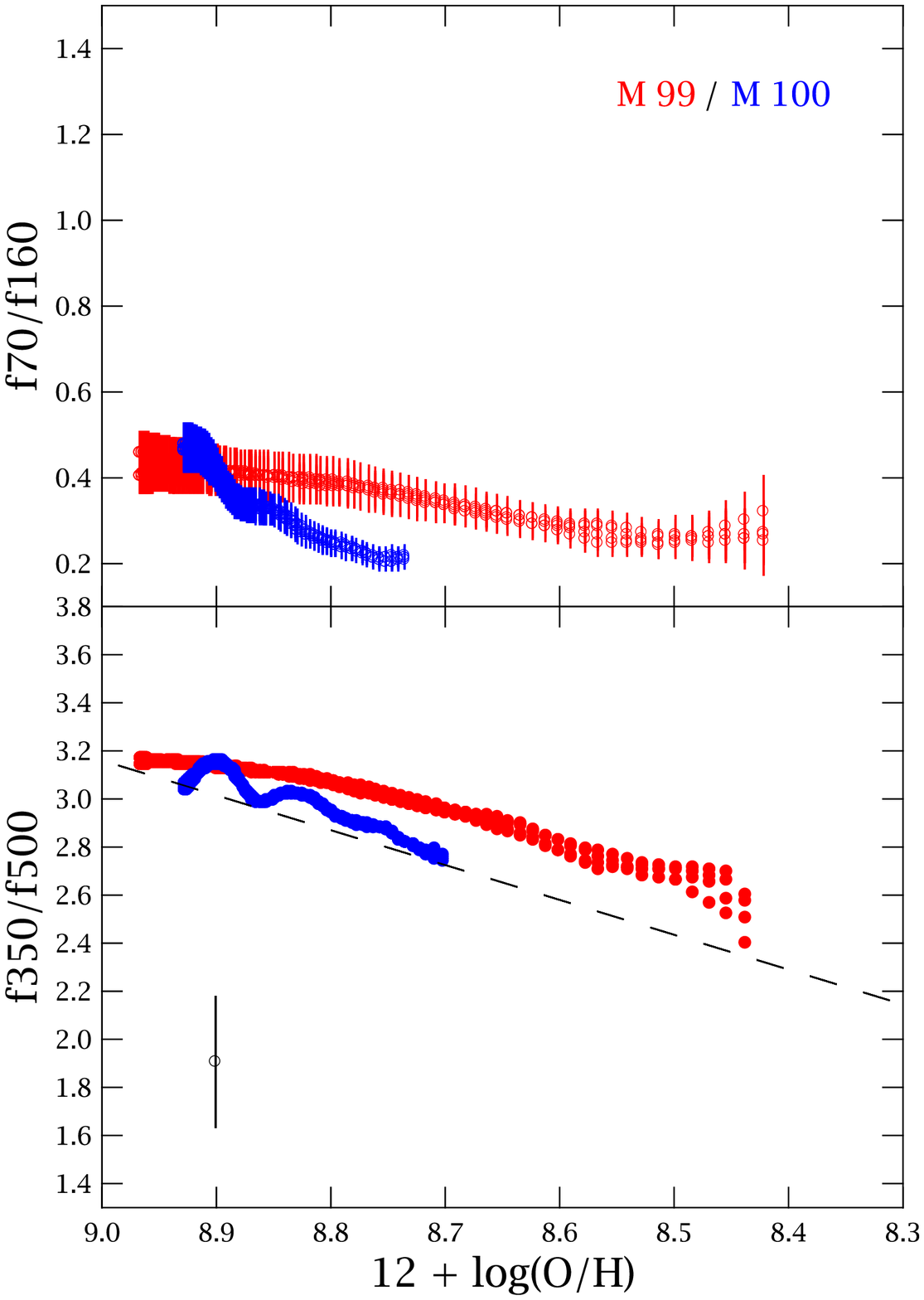}
\end{center}
\vspace*{-0.3cm}
\caption{{\it Left:} Radial surface brightness profiles for M\,99 {\sl(top)} and
  M\,100 {\sl(bottom)} obtained from the smoothed and matched maps
    (see text). The profiles are from the bottom up: MIPS 24, 70, 160
  $\mu$m, SPIRE 250, 350, 500 $\mu$m (all in [$\log({\rm Jy}/{\rm
      arcsec}^{2})$]), \hi\ {\sl(black)} and {\bf \hii}${^5}$
   ({\sl gray}, both in [$\log({\rm M_{\odot}}/{\rm pc}^{2})$]) adding 0.5, 0,
  0, 1, 2, 3, -1, -1, respectively, to separate them in the plot and the
  radial axis is in units of $R25$. In all cases, we plot four different radial
  profiles per galaxy to show the influence of the fixed ellipse fitting (see
  \sec\ref{analysis}). {\sl Middle:} SPIRE colours $f350/f500$ and $f250/f350$
  with radius for M\,99 {\sl(red)} and M\,100 {\sl(blue)}. {\sl Right:}
  $f350/f500$ and $f70/f160$ versus the metallicity again for M\,99 in {\sl
    red} and M\,100 in {\sl blue }. Since the uncertainty in the SPIRE flux is
  still rather large, we only provide an indication of the error in the SPIRE
  colours with a {\sl black open circle} in the lower left corner of the
  appropriate panels (see \sec\ref{analysis}). The {\sl dashed black line} in
  the lower right panel is a fit to the integrated properties of a larger
  sample of galaxies with a variety of morphologies \citep{Boselli2010b}. }
\label{profs}
\vspace*{-0.2cm}
\end{figure*}

The 24, 70, and 160 $\mu$m images were part of the SINGS survey 
\citep{Kennicutt2003} and were processed using the MIPS Data Analysis Tools
\citep{Gordon2005} along with techniques described by \cite{Bendo2010a} and
Clements et al. (2010, in preparation). The FWHM of the MIPS beams are about
6\arcsec, 18\arcsec, and 38\arcsec, the pixel sizes are 1.5\arcsec,
4.5\arcsec, and 9\arcsec\ per pixel at 24, 70, and 160 $\mu$m, respectively.
The CO($J$=1-0) maps, used as the tracer of the molecular hydrogen dominating
the molecular mass, are taken from the Nobeyama CO Atlas of Nearby Spiral
Galaxies \citep{Kuno2007}. The FWHM is 15\arcsec\ and the pixels are
1\arcsec\ per pixel.  For the \hi, we used the zeroth moment maps from the VIVA
survey \citep[VLA Imaging of Virgo Spirals in Atomic Gas;][]{Chung2009}. The
FWHM and pixel sizes are $\approx\!37$\arcsec\ and 5\arcsec\ per pixel for
M\,99 and $\approx\!30$\arcsec\ and 10\arcsec\ per pixel for M\,100.

\subsection{Analysis}
\label{analysis}

The SPIRE maps were first converted into Jy per pixel assuming a Gaussian beam
(of the above quoted sizes). For the \hi, we used the elliptical beam sizes
given in \cite{Chung2009}. In all maps, we masked strong sources and
artefacts.  However, in the case of the SPIRE maps, being confusion limited,
all faint sources are unmasked and part of the background. The residual
background on each map was subsequently determined using IRAF's\footnote{Image
  Reduction and Analysis Facility
  (IRAF) \newline \hspace*{0.5cm}http://iraf.noao.edu/} \textsc{ellipse} task
as described in \cite{Pohlen2006}. The background-subtracted maps of all
wavelengths were thereafter smoothed to the MIPS 160$\mu$m resolution of
$\approx 40$\arcsec\ using custom convolution kernels derived as described in
\cite{Bendo2010a} and then matched to the SPIRE 500$\mu$m pixel size of
14\arcsec.

The SPIRE 250$\mu$m map was chosen to derive the final set of ellipse-fitting
parameters (i.e., ellipticity, position angle, and centre). To ensure that our
results are independent of the particular ellipse geometry selected, we
applied four different fixed ellipse fits to each map. For example, M\,99
being a one-armed spiral, is slightly asymmetric (\cf\fig\ref{maps}) so we
selected one set of ellipse parameters derived in the outer parts and one in
the inner parts, which each have slightly different centres. The final
radial profiles, obtained using a combined mask on the smoothed and matched
maps, are shown in \fig\ref{profs} out to where we can trace signal on
the map. The error bar in each measured point is a combination in quadrature
of the uncertainty in the overall absolute calibration (currently for SPIRE
the dominating source), the error in the ellipse intensity from the
\textsc{ellipse} task, the uncertainty in the estimate of the background, and
an additional uncertainty calculated by comparing the results from
different versions of the pipeline. This last, very conservative uncertainty, is
responsible for the currently rather large error bars in the measured flux
ratios in \fig\ref{profs}.\\[-0.25cm]

\section{Results}

We detect dust emission traced by all three SPIRE bands out to at least
the optical radius defined by $R25$ for both galaxies (see \fig\ref{maps} for
$\approx\!3\!-\!5 \sigma$ detections in the nominal maps, and \fig\ref{profs}
for the deeper, radially averaged profiles from the smoothed maps). Compared
to \hi, the dust can be found almost out to the \hi-edge of the regular disk
for M\,100. This is not entirely surprising, since M\,100 is an intermediate
\hi-deficient \citep{Haynes1984, Cayatte1994} galaxy (\hi-def$=\!0.35$) and
thus its outer \hi-disk has probably been already stripped by the interaction
with the cluster environment \citep{Boselli2006}. A similar extension of the
dust and \hi-disk is observed for this range of \hi\ deficiencies in other
Virgo cluster galaxies \citep{Cortese2010}.
The situation is different for M\,99, which is not \hi-deficient
(\hi-def$=\!-0.1$). Figure \ref{maps} clearly shows that the \hi\ emission is
much more extended than the submm at least to the north. Interestingly, this
extended \hi\ halo \citep{Chung2009} however might be barely detected, but at
the moment we cannot exclude that this is caused by residual background
inhomogeneities coupled with a cluster of unresolved background sources.  We
can however exclude the presence of submm emission corresponding to the giant
\hi\ tail of M\,99 \citep{Haynes2007} to the southwest and we also find no
measured flux associated with the extended low surface brightness feature to
the southwest of M\,100 \citep{Chung2009}.

The left column in Figure \ref{profs} shows the radial profiles obtained by
the ellipse fitting to the smoothed maps. The MIPS and \hi\ profiles agree
with the published ones by \cite{Munoz2009b} and \cite{Chung2009},
respectively. For M\,99, the MIPS and SPIRE profiles follow similar trends
including a weak radial break in the profile, i.e., a change in the slope, at
$\approx\!0.6\times R25$, and a broken exponential is a more accurate fit than
a single exponential \citep[e.g.~][for more background on
  breaks]{Pohlen2006}. This break is also visible in the optical profile shown
by \cite{Munoz2009b}. The same is true for M\,100, which also exhibits a more
obvious break at around the same distance \citep[it is even more striking in
  the profile at native resolution presented by][]{Sauvage2010}.
This is the first time we see these breaks clearly in the dust distribution,
while they are well-known at optical wavelengths. None of the so far presented
hypotheses for the origin of these breaks have addressed this before
\citep[see e.g.~the recent review by][for references]{Vlajic2010} and it will
be another pice of the puzzle to be explained by the various proposed
models. The rising profile in the inner part of M\,100 is related to the more
prominent bulge, bar, or inner-disk component, which is discussed in more
detail in \cite{Sauvage2010}.

To investigate the variation in submm colours as a function of
radius we plot in the middle column of Figure \ref{profs} the ratio of the
SPIRE bands $f350/f500$ to $f250/f350$. These are colour temperature
indices. The advantage of using these instead of a derived dust mass is that
they are independent of the specific, not yet well studied, model assumptions
in this new wavelength range. They both decrease with radius, which suggests
that the dust in the outermost regions is colder than in the centre of the
galaxies. This is naturally explained by an interstellar radiation field
becoming less intense in the outskirts.
Our profiles are very similar to those of M\,81 presented in
\cite{Bendo2010b}, who argue that the radial variation is driven by heating
from the evolved stars in the galaxy. The observed range of flux ratios along
the galactic radii is the same as found for a sample of galaxies with a
wide variety of morphologies using integrated SPIRE fluxes
\citep{Boselli2010b}. Interestingly, the agreement between their
integrated and our resolved analysis extends beyond the colour profiles
as shown in the right panels of \fig\ref{profs}, where we couple our colour
gradients to the metallicity gradient published by \cite{Skillman1996}
renormalised to the [OIII]/[NII] base of \cite{Pettini2004}. The trend we
observe radially for M\,99 and M\,100 matches the fit to the integrated
properties of the \cite{Boselli2010b} sample very well. Both $f350/f500$ and
$f70/f160$ (albeit only very weakly) decrease with the radially decreasing
metallicity. This is again expected since a lower activity of star formation
in the outer parts, as observed by \cite{Wilson2009}, consequently entails lower
metallicities.

In \fig\ref{gas2dust} we finally show the ratio of the total gas mass
(\hi\ plus \hii)\footnote{We use a radially constant X-factor as given in
  \cite{Kuno2007}} to 500$\mu$m flux ratio for the two galaxies. Since the
500$\mu$m flux is a proxy of the dust mass, this provides a
"model-independent" indication of the radial evolution of the dust-to-gas
ratio. There is a clear trend visible with the gas-to-dust ratio
  increasing radially, which is consistent with earlier results
\citep{Munoz2009a, Bendo2010a}, but the exact shape needs to be revised once a
proper SED dust modelling including the new SPIRE bands is available.

In conclusion, we have found that the dust emission can be traced by the SPIRE
bands at least out to the optical radius and beyond. The dust shows the same
breaks in the radial profile as seen in the optical. The \hi is only slightly
more extended but this needs to be regarded here in the context of the cluster
environment. The SPIRE colour temperature indices decrease with radius
following the measured trends in metallicity, and the extent of the measured
values along the galaxies' radii is consistent with the integrated properties
of galaxies with a variety of morphologies. We have shown evidence of a
radially rising gas-to-dust ratio. These results provide the first indication
of the improved capabilities {\sl Herschel} can offer for studying the
resolved dust distribution in galaxies.

\begin{figure}
\begin{center}
\includegraphics[width=6.65cm,angle=0]{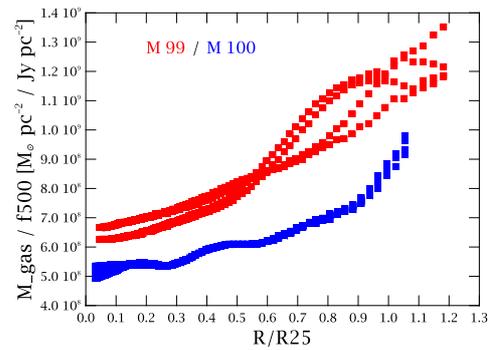}
\end{center}
\vspace*{-0.4cm}
\caption{Total gas mass (\hi\ plus \hii) to SPIRE 500$\mu$m flux ratio.}
\label{gas2dust}
\vspace*{-0.1cm}
\end{figure}


\begin{acknowledgements}
SPIRE has been developed by a consortium of institutes led by
Cardiff Univ. (UK) and including Univ. Lethbridge (Canada);
NAOC (China); CEA, LAM (France); IFSI, Univ. Padua (Italy);
IAC (Spain); Stockholm Observatory (Sweden); Imperial College
London, RAL, UCL-MSSL, UKATC, Univ. Sussex (UK); Caltech, JPL,
NHSC, Univ. Colorado (USA). This development has been supported
by national funding agencies: CSA (Canada); NAOC (China); CEA,
CNES, CNRS (France); ASI (Italy); MCINN (Spain); SNSB (Sweden);
STFC (UK); and NASA (USA). Thanks to Tom Hughes for providing the recalibrated
metallicities. The SDSS jpg was taken from http://www.sdss.org using the
Finding Chart tool. \\[-0.7cm]
\end{acknowledgements}

\end{document}